# Evaluation of Microencapsulation of The UFV-AREG1 Bacteriophage in Alginate-Ca Microcapsules using Microfluidic Devices

Delaine M. G. Boggione[a*], Laís S. Batalha[a], Marco T. P. Gontijo[a], Maryoris E. S. Lopez[b], Alvaro V. N. C. Teixeira[c], Igor J. B. Santos[d], Regina C. S. Mendonça[a]

*a – Departamento de Tecnologia de Alimentos, Universidade Federal de Viçosa, Viçosa, Minas Gerais, Brasil.*
*b – Departamento de Ingeniería de Alimentos, Universidade de Córdoba, Colombia.*
*c – Departamento de Física, Universidade Federal de Viçosa, Brasil.*
*d - Departamento de Química, Biotecnologia e Engenharia de Bioprocessos, Universidade Federal de São João Del Rey, Campus Alto Paraopeba – Ouro Branco, Brasil.*

ABSTRACT

The indiscriminate use of antibiotics and the emergence of resistant microorganisms have become a major challenge for the food industry. The purpose of this work was to microencapsulate the bacteriophage UFV-AREG1 in a calcium alginate matrix using microfluidic devices and to study the viability and efficiency of retention. The microcapsules were added to gel of propylene glycol for use as an antimicrobial in the food industry. The technique showed the number of the phage encapsulation, yielding drops with an average 100-250 μm of diameter, 82.1 ± 2% retention efficiency and stability in the gel matrix for 21 days. The gel added to the microencapsulated phage showed efficiency (not detectable on the surface) in reducing bacterial contamination on the surface at a similar level to antimicrobial chemicals (alcohol 70%). Therefore, it was possible to microencapsulate bacteriophages in alginate-Ca and apply the microcapsules in gels for use as sanitizers in the food industry.

## 1. Introduction

The indiscriminate use of antibiotics and the emergence of resistant microorganisms have aroused interest in alternative methodologies to control pathogens and spoilage microbes in foods and in different areas of the food industry. Among these methodologies, the use of bacteriophages, bacteriocins and essential oils, which can be widely used in food packaging, stands out [1, 2].

Recently, many studies have examined the most effective processes to reduce or eliminate pathogenic microorganisms that appear in commodities of animal origin. These processes include the implementation of food safety systems, good agricultural and manufacturing practices, consistent and continuous training, and multidisciplinary research. Among these pathogens, *Salmonella* spp, *Listeria monocytogenes* and *Escherichia coli* O157:H7 are the most important. *E. coli* O157:H7 has been associated with outbreaks involving lettuce, spinach, culinary herbs and other fresh products [3]. According to Shekarforoush and co-workers [4] *E. coli* is considered an important agent which can spread through direct contact and by consumption of contaminated foodstuff, surviving to a 15% concentration of acetic, citric and lactic acid. This bacterium can cause hemorrhagic colitis, hemolytic uremic syndrome (HUS), thrombocytopenic purpura and, occasionally, death.

The use of bacteriophages has become a promising therapy, especially due to bacterial resistance to antibiotics, which represents an important factor in public health due to the mortality caused by multiresistant bacteria. Furthermore, it also decreases concerns about antibiotic residues in foods and the compulsory retention period by humans [5]. The main applications of bacteriophages have been by oral administration of solutions containing the phages in therapy, or by spraying foods or equipment surfaces for decontamination [6, 7]. Another emergent technique is the application of encapsulated bacteriophages.

Essentially, encapsulation consists of a process in which the material to be encapsulated is maintained inside the encapsulating material to minimize lesions or losses. These losses can occur during passage through the gastrointestinal tract, during industrial processes and others, which can minimize the antimicrobial effect of bacteriophages. The encapsulation also permits the controlled liberation of viral particles under the influence of specific conditions [8].

*Corresponding author.:
E-mail address*: delaine.gouvea@ufv.br

In this context, the methods and equipment of microfluidics can be used for the encapsulation of micro and nanostructures in polymeric microgels [9, 10]. The system is constituted by channels ranging from tens to hundreds of micrometers, with different geometries and the most varied materials, such as polymers, glass, ceramics, metals, etc. [10, 11]. The manipulation of multiple-phase flow in the device allows low polydisperse emulsions to be generated, which may be single or double emulsions, of several diameters, expanding the fields of application. The higher degree of control offered by this method and the completely distinct fluid streams make microfluidics a very promising and versatile technique [12]. The microfluidic device can be utilized to produce microcapsules, microparticles and even nanoparticles, depending on the manufacture and design of the equipment [10].

Polymers that are purchased from natural sources of renewable resources are even more interesting because they have a low cost, water solubility, biocompatibility, biodegradability, ability to gel and good availability. Sodium alginate is a natural nonbranched polysaccharide (extracted from brown seaweed and some bacteria) formed by two types of polyuronic acid, β-D-mannuronic acid (M) and α-L-guluronic acid (G) bound by (1 → 4)-glycosidic bonds, of varied composition and sequence. These monomers are epimers with different orientations in the polymer chain, and the G unit is responsible for allowing cross-ionic bonding. The gel is formed by connecting the units of guluronic acid with cations, resulting in a three-dimensional network that is maintained by ionic interactions [13-16].

Sodium alginate is widely used as a gelling agent due to its ability to form gels with divalent cations, such as $Ca^{+2}$, $Ba^{+2}$ or $Sr^{+2}$. The use of alginate is appropriate for encapsulation owing to its GRAS (Generally Recognized As Safe) aspect and absence of toxicity [17]. Because of the relatively mild gelling process, Ca-alginate particles have found numerous applications as encapsulation vehicles and delivery systems for cells, proteins, enzymes and drugs, as well as useful tools for the controlled release of drugs in biomedical science and pharmaceutical engineering [18].

Based on these premises, the aim of this work was to microencapsulate the bacteriophage UFV-AREG1 in a matrix of calcium alginate, using microfluidic devices. The microcapsules produced by the device were added to a propylene glycol gel for application in the sanitization of food surfaces, evaluating the antimicrobial effect. To our knowledge, there is no work in the literature about the encapsulation of bacteriophages using microfluidics for possible applications in the food industry; thus, this is an initiative to increase phage stability and viability during storage.

## 2. Material and methods

### 2.1 Microorganism

The bacterium Escherichia coli O157:H7 (ATCC 43895) was acquired from the Fiocruz (Fundação Oswaldo Cruz) collection. The bacteriophage UFV-AREG1, isolated from cowshed wastewater, showed specificity for enterohemorrhagic *E. coli* O157:H7 (ATCC 43895), *E. coli* 0111 (CDC 011ab) and *E. coli* (ATCC 23229). The genome sequence of the bacteriophage UFV-AREG1 is available under the accession number KX009778.3 in GenBank [19].

### 2.2. Determination of the diameter of viral particles

The dynamic light scattering technique is a method of investigation in the macromolecular system, characterized by being a noninvasive method and allowing small volumes to be used.

The technique was used to determine the average hydrodynamic radius (HR) and size distribution of viral particles. The equipment used was Dynamic Light Scattering (DLS), Static Light Scattering (SLS) (Brookhaven Co. - USA) and light scattering Zetasizer (Zetasizer Nano S – Malvern – UK). In order to obtain homogeneous samples, a filtration was made using syringes with a hydrophilic 13 mm cellulose acetate membrane (Analitica, USA) with 0.22 μm pore size. All measures were made under controlled temperature (25.0 ± 0.5 ºC). The measures of light scattering were performed at angles ranging from 30 ° to 135 °, at intervals of 15 °. The technique is appropriate for determination of structures of 10 to 1000 nm. The second-order cumulant model was adjusted using the data, from which was obtained the average relaxation rate of decay of this correlation ($\bar{\Gamma}$). The measurement was done three times, and the average diameter of particles was reported by the linear adjustment of a graph $\bar{\Gamma}$ *versus* $q^2$, where $q$ represents the scattering vector module. The slope obtained by this adjustment is self-diffusion coefficient, $D$, of the structures. The hydrodynamic radius was calculated using the Stokes-Einstein equation.

### 2.3. Microfluidic device

The device function was to produce emulsions with low-size dispersion from immiscible liquids.

Two cylindrical capillary tubes were used. The hydrophobic treatment was applied with 10% (v/v) solution of octadecyltrimethoxysilane in toluene. To mount a cylindrical device, one was placed inside the other. The dispersion of liquids in the device was controlled by the sheer force to release the drops, producing water/oil emulsion (w/o) as shown in Figure 1. The connections were made to set the entries of the dispersed and continuous liquid phase in the device, which were placed in syringes; the region where the emulsion formed is called constriction. The syringes were connected to syringe pumps (PHD 4400 Programmable - Harvard Apparatus, USA), which controlled the entry of the liquid. The experiment was monitored by an optical microscope, which was connected to an image-capture CMOS camera connected to the computer.

### 2.4. Production of calcium alginate microcapsules

For the calcium alginate matrix, 1.5 % (w/v) of sodium alginate (Vetec, Brazil) was added to purified water. The mixture was homogenised in a magnetic stirrer until complete dissolution of alginate. The phage solution 106 PFU·mL$^{-1}$ (1:10 v/v) was added to the mixture. After this procedure, the solution was filtered through nylon filters (Analitica, USA) of 0.45 μm within the syringe, and this was the dispersed phase. Sunflower oil with SPAN 80 (Sigma-Aldrich, USA) (0.3% w/v) added was also filtered, and this was the continuous phase. The syringes were coupled to the flow pump and the microfluidic device. Flow rates were controlled to verify the formation and detachment of droplets in a regular manner, and to determine the desired size of the microcapsules. For the dispersed phase 50-100 $\mu L \cdot h^{-1}$ was used, and 1000-1500 $\mu L \cdot h^{-1}$ for the continuous phase. The droplets produced were collected in a $CaCl_2$ solution (Vetec, Brazil) (0.1 $mol \cdot L^{-1}$) to form the alginate-Ca microcapsules (Figure 1). After a few minutes of cation exchange (approximately 20 min), alginate-Ca microcapsules were obtained.

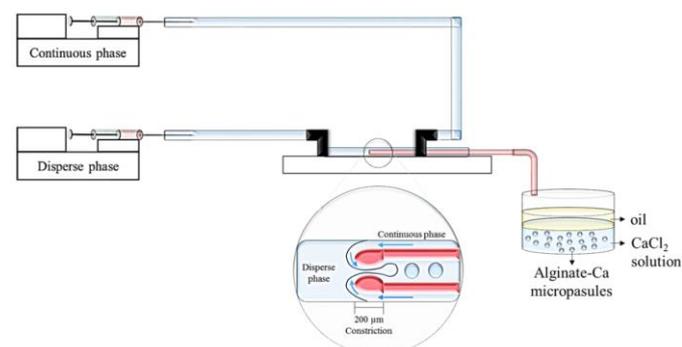

**Figure 1:** Microfluidic device and Alginate-Ca microcapsule production.

After formation of the microcapsules, the excess oil and the $CaCl_2$ solution were removed with Pasteur pipette and the microcapsules washed with water to remove oil residues. The water was withdrawn with the pipette, leaving only the microcapsules, which were stored in 1.5 mL microtubes and reserved in refrigeration temperature for evaluation of retention efficiency and subsequent addition in propylene glycol gel.

### 2.5. Retention Efficiency (RE) for the Ca-alginate matrix and stability of microcapsules in propylene glycol gel

For retention efficiency of the bacteriophages in Ca-alginate the titration of microencapsulated bacteriophages was determined. The dissolution of the microcapsules for releasing the bacteriophage was performed as described by Liu and co-workers [20]. The microcapsules were suspended in Broken-Microsphere Solution (MBS) containing sodium citrate (Neon, Brazil) 50 $mmol \cdot L^{-1}$, sodium bicarbonate (Neon, Brazil) 0.2 $mol \cdot L^{-1}$ and Tris-HCl (Sigma-Aldrich, USA) 50 $mmol \cdot L^{-1}$, pH 7.5 for a few minutes until total dissolution of the microcapsules for phage release. The bacteriophages released from microcapsules were serially diluted using SM buffer (50 $mmol \cdot L^{-1}$ Tris-HCl - Sigma-Aldrich, USA [pH 7.5], 0.1 $mol \cdot L^{-1}$ NaCl - Vetec, Brazil, 8 $mmol \cdot L^{-1}$ $MgSO_4 \cdot 7H_2O$ - Chemco, Brazil, 0.01% gelatin - Merck, USA) to obtain from 10 to 100 plaque-forming units (PFU). On a plate containing agar BHI base (Brain Heart Infusion - Himedia, India + agar 1.2% - Himedia, India) 100 mL of bacterial culture in log phase and 100 μL of phage suspension dilution were added to a tube containing agar overlay (Brain Heart Infusion broth - Himedia, India + agar 0.6% - Himedia, India) and poured on the plates. The plates were incubated at 35 ± 2 °C / 18- 24 h.

The retention efficiency (RE) of bacteriophages in alginate microcapsules was determined by equation 1 per Gomes and co-workers [21], adapted to bacteriophages.

$$RE = \frac{\text{entrapped quantity of bacteriophages}}{\text{original quantity of bacteriophages}} \times 100 \quad \text{Eq. 1}$$

Where "entrapped quantity of bacteriophages" is the concentration of bacteriophages present in the microcapsules and the "original quantity of bacteriophages" indicates the concentration of bacteriophages initially used to make the microcapsules of alginate.

The stability was assessed for 21 days (0, 7, 14 and 21 days). To determine the stability of microcapsules added to the propylene glycol gel, the titration proceeded as described above. After incubation for 18-24 h / 37 °C, the lysis plaques were counted and expressed as plate forming units ($PFU \cdot mL^{-1}$).

### 2.6. Characterization of microcapsules containing bacteriophages by confocal microscopy

The characterization was performed to assess the homogeneity of phage solution within the microparticles and stability. Before encapsulation, the UFV-AREG1 phage was labeled with fluorescein isothiocyanate, fluorescence emitted at 488 nm (FITC, Sigma-Aldrich, USA). The UFV-AREG1 bacteriophage was labeled with isothiocyanate using the procedure described by Bakhshayeshi and co-workers with modifications [22]. Briefly, 1.2 ml of phage suspension in concentration of $10^6$ $PFU \cdot mL^{-1}$ in 0.1 M borate buffer pH 9.2 was mixed with 0.021 g of FITC and 5 mL of N,N-dimethylmethanamide (also known as dimethylformamide or DMF - Sigma-Aldrich, USA). The resulting mixture was left overnight at 4 °C. Dialysis of the mixture was carried out to remove excess of fluorophore. The buffer was changed every four hours.

Next, the same phage sample was encapsulated per the methodology described in Section 2.4 and visualized by confocal microscopy (Confocal Microscope - LSM 510 Meta Laser Scanning - Zeiss, Germany) at the Núcleo de Microscopia e Microanálise, Universidade Federal de Viçosa. Through fluorescence radiation of the marked phages the homogeneity of the distribution of the phage around the microcapsules was verified.

### 2.7. Application of Ca-alginate microcapsules in a propylene glycol gel and antimicrobial evaluation

The gel was prepared using the following ingredients: 2% (w/v) hydroxymethyl cellulose (Mapric, Brazil); 5% (w/v) propylene glycol (Mapric, Brazil); 0.1% (w/v) EDTA (Mapric, Brazil); 0.2% (w/v) methylparaben (Mapric, Brazil). The volume was completed to 100% with deionized water. With the water under heating and stirring the hydroxymethyl cellulose was sprayed, then propylene glycol, EDTA, and methyl paraben were added. The system was kept under stirring until a temperature of 70 °C. After reaching that temperature, the heat was turned off. Under stirring, the microtube (1.5 mL) containing alginate microcapsules with phage, was added and mixed in solution. The solution was stored in appropriate containers for further use. Alcohol Gel 70% purchased commercially was also used as an antimicrobial in comparison to the manufactured gel and a control surface containing only bacteria, without the use of antimicrobials.

To evaluate the stability of the microcapsules in the gel, to a tube containing 1.0 mL of gel with microcapsules, more phage was added to 4.0 mL of MBS solution for total dissolution and release of phage. From this stock solution, 1.0 mL was transferred to a tube containing 4.0 mL of SM buffer, proceeding to serial dilution. To evaluate the antimicrobial efficiency of the gel with microcapsules, alcohol gel and surface control, microbiological analyses were performed in vitro using *E. coli* O157:H7. The pre-activated bacteria at a concentration of $10^4$ $CFU \cdot mL^{-1}$ were added to a previously sterilized laboratory bench area divided into three parts, allowing the bacteria to act for 20 minutes, which is the average generation time of *E. coli*. Next, to first part gel with microcapsules was applied, to the second part alcohol gel 70%, and to the third nothing was applied. The swab technique is considered a standard methodology for microbiological analysis by the American Public Health Association (APHA).

The swab technique is used to evaluate surfaces and handlers' hands in the food industry. Sterilized swabs were moistened in tubes containing 10 mL of 0.85% saline solution and rubbed over the analyzed areas. After collection, the part touching the swab shaft was broken on the inside of the tube edge, plunging the swab with material collected into saline 0.85%. This was homogenized and decimal dilutions proceeded. The diluent was then analyzed by 0.1 mL aliquots plated on agar plates in EMB (Eosin Methylene Blue - Himedia, India) selective for *E. coli*. The EMB plates were incubated at 37 °C ± 2 °C / 24-48 h. The results were expressed as CFU/unit. The control surfaces without the use of antimicrobial agents were also assessed.

*2.8. Statistical analysis*

For alginate-Ca matrix, a completely randomized design was adopted with three replications. The results were submitted to analysis of variance (ANOVA) with F test. When there was a significant difference in the F test at 5% probability, the Tukey test was applied. The results were evaluated using the Minitab 16® program.

## 3. Results and discussion

*3.1. Dynamic light scattering (DLS)*

Dynamic light scattering was used to characterize the phage particles. The average diameter of the particles was $107 \pm 1$ nm ($\tau$) that was calculated by the usual Stokes-Einstein equation from the diffusion coefficient obtained by the linear adjustment of a graph $\bar{\Gamma}$ versus $q^2$. This $\tau$, called hydrodynamic diameter, is the diameter of the sphere with the same diffusion coefficient. This result was confirmed by transmission electronic microscopy (Figure 2), which showed the icosahedral head (114 by 86 nm) and contractile tail (117 by 23 nm) of the bacteriophage. These morphological parameters allowed the UFV-AREG1 phage to be included in the Caudovirales order and Myoviridae family [19].

*3.2. Microfluidic device and production of alginate-Ca microcapsules*

The size of the microcapsules was determined using the ImageJ program, and the scale used to calculate the size was 1.7513 micrometers/pixel for the images. Initially, we opened the photograph corresponding to the desired flows; then we drew a square in which the drop should be circumscribed. The side of the square should have the same length as the diameter. We did this for several drops and obtained the mean value of the diameter and the standard deviation of the same. The average diameter of the alginate-Ca microcapsules ranged from 100 to 250 μm (Figure 3). By statistical data, the size of the alginate particles of Figure 3G was calculated, where about 40 particles were analyzed corresponding to a diameter of 251 μm with a standard deviation of 24.3% (Figure 4), showing that the system is not homogeneous.

However, according to their size, the capsules are classified as nanoparticles or microparticles, ranging from (0.01 to 0.2) nm and (1 to 100) μm [23], and our microcapsules are in the order of micrometers and can be considered satisfactory for applications in antimicrobial gels, in which the microcapsules are added and homogenized in the solution in an unnoticeable visibility. As the gel is for cleaning the hands/surfaces, larger microcapsules can generate a sensation that would be unpleasant on the hands of manipulators, for example.

Encapsulation using microfluidic devices allows the production of low polydisperse emulsions in this work, which are formed by a controlled flow. The device and the formation of the capsules can be observed in Figure 3 (A - G). By optical microscopy it was possible to observe the droplet formation coming out of constriction, with a circular shape and with homogeneous diameters in a flow of 50 μL·h$^{-1}$ of the dispersed phase and 1000 μL·h$^{-1}$ of the continuous phase; however, by the end of the capillary to be collected droplets have undergone coalescence and increased in size. This justifies the size variation when falling into the CaCl2 solution. Preliminary tests were performed to achieve the appropriate flow combination. In higher flow rates of the dispersed phase (200 µL·h$^{-1}$ and 1500 µL·h$^{-1}$, respectively) there is the formation of larger microcapsules, as shown in Figure 3 (E-G). Variations in flow rates demonstrate that there can be difference in the average diameter of the droplets.

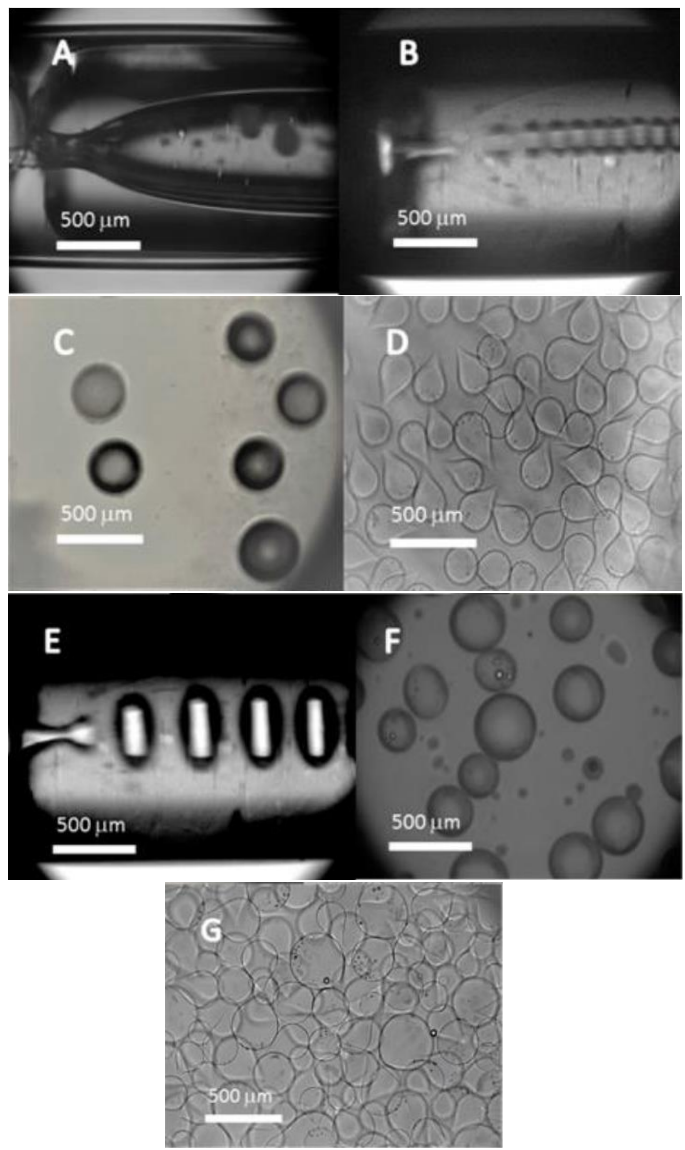

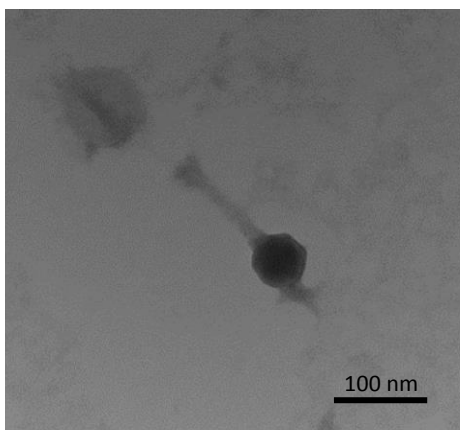

**Figure 2:** Electron micrograph of phage UFV-AREG1.

**Figure 3:** Microfluidic device and production of microcapsules with a flow rate of (50 μL·h$^{-1}$ - 200 μL·h$^{-1}$) and (1000 μL·h$^{-1}$ - 1500 μL·h$^{-1}$) for the dispersed and continuous phase, respectively. (A) device; (B) and (E) droplet formation in the constriction; (C) and (F) formed droplets alginate-Na with phage before contact with the CaCl$_2$ solution; and (D) and (G) alginate-Ca microcapsules containing the phage after contact with the CaCl$_2$ solution.

The droplets of sodium alginate were collected in a 0.1 M solution of $CaCl_2$ (Figure 1D), forming the calcium alginate microcapsules containing the microencapsulated bacteriophage. Once in contact with the chloride ions in solution, the capsules did not remain in a homogeneous size and, due to the impact of the fall, the capsules were misshapen (Figures 3D-G). The distance from the outlet capillary to the collection of drops in $CaCl_2$ solution may have caused the deformation and the size difference of the microcapsule, because the drops in the capillary outlet pooled, and then fell into the chloride solution. These conditions may have occurred due to the lack of preparation in the use of the equipment.

Hu and co-workers report in their work that several types of alginate microgels, such as hemi-spherical, mushroom-like, disk-like, and red blood cell-like, among others, were prepared by adjusting the flow rate of the microfluidics and the gelation conditions. Besides flow rate, a unique external ionic crosslinking (gelation bath) was designed. The morphology of the microgel can be finely tuned continuously by simply varying the gelation conditions. A key advantage of this procedure for preparing microgels is that the morphology of the alginate microgels can be tuned continuously by simply changing height, as can the viscosity of the gelation bath by varying the content of glycerol and owing to the interplay between the viscous and elastic forces during the subsequent gelation process [24].

Hu and co-workers employed simple capillary devices to evaluate the effect of flow rate and the impact on the formation of alginate-Na capsules, which formed nonspherical microcapsules after falling into the surfactant solution. This morphology is influenced by the contact speed, the droplet size and the concentration of the alginate solution. They concluded that depending on the concentration of the alginate solution, the droplets presented different rheological properties and, thus, showed different morphologies [25].

C.-H. Yeh [26] using T junction microfluidic devices established flow rates of 0.05 mL·min$^{-1}$ and 0.5 mL·min$^{-1}$ for the dispersed and continuous phase, respectively, in the microencapsulation of nanoparticles in an alginate matrix, producing emulsions of 200 ± 10 µm diameter homogeneously distributed, and they verified that increasing sample flow rate (under fixed oil flow rate) contributes to an increase in the emulsion sizes, and that increasing oil flow rates leads to a decrease in the emulsion sizes. The work developed by Nisisako and contributors [27], also in a T junction device, produced droplets of 100 µm, increasing the flow rate of the continuous phase. Increasing the flow of the continuous phase the droplet size diminished. Changes in the flow rate also established the number of droplets produced by the device. Maintaining the dispersed phase flow constant and varying the continuous phase flow, a higher or lower production of droplets was obtained. These authors showed that the droplet production rate is a function of the flow of the continuous phase and, consequently, the production rate can be linearly increased.

Huang and contributors [28] proposed an in situ method of alginate microcapsule production of controlled diameter, which can be an intelligent delivery system. The method resulted in two different diameters of alginate microcapsule produced and separated in microfluidic devices. The particles showed relative size homogeneity. They also noted that, in both particles, the size could be changed. Their results showed that the size droplets are equivalent to the channel diameter, and can be adjusted by varying the flow speed of the continuous phase in relation to the dispersed phase. Since the separation efficiency is high in this system, both droplet sizes are monodispersed and have elevated reproductivity. Ren and contributors [29] also developed a microfluidic device to produce monodispersed alginate microcapsules with double emulsion for the encapsulation of pharmaceuticals and active chemical products. The average capsule diameter was 117 and 255 µm for the internal and external diameters, respectively. Our work showed that there was variation of the particle size when they were collected in the $CaCl_2$ solution. The size can be influenced by the coalescence occurring at the end of the collection tube and the deformity by the height of the tube until the solution influenced by the rheological properties of the sample analyzed. The parameters and careful handling of the equipment need to be adjusted so that the formation of the alginate-Ca microcapsules maintains a low deformity and low size dispersion.

### 3.3. Retention Efficiency (RE)

The average initial concentration of the bacteriophage (expressed in PFU·mL$^{-1}$) added to the solution of sodium alginate-Na 1.5% before injecting into the microfluidic device was $2.0 \times 10^6$ PFU·mL$^{-1}$. The final concentration of the phage retained in the capsules was $1.4 \times 10^5$ PFU·mL$^{-1}$. Therefore, the retention efficiency using the microfluidic device to encapsulate the UFV-AREG1 in the calcium alginate microcapsules was 82.1% ± 2%. Huang and co-workers proposed a device to microencapsulate chondrocyte cells, showing efficiency of 94 ± 2% viability of the cell after encapsulation. The device was capable of producing alginate microspheres or microencapsulating in a simple, continuous, controllable and homogeneous way, showing less contamination [30].

Liu and co-workers, using a microfluidic device to produce double emulsion of calcium alginate microcapsule, obtained a diameter of 323 µm and 227 µm for the external and internal emulsions, respectively. The same authors used bovine serum albumin protein as a model to encapsulate, and they obtained positive results for the encapsulation, determining that the methodology is viable for the encapsulation and to carry drugs and active chemical products [31].

In another study using flow-focusing microfluidic devices to microencapsulate ascorbic acid in solid lipidic microcapsules to develop fortified foods, the retention efficiency in the lipidic matrix ranged from 74 to 96% per the initial composition on the matrix.

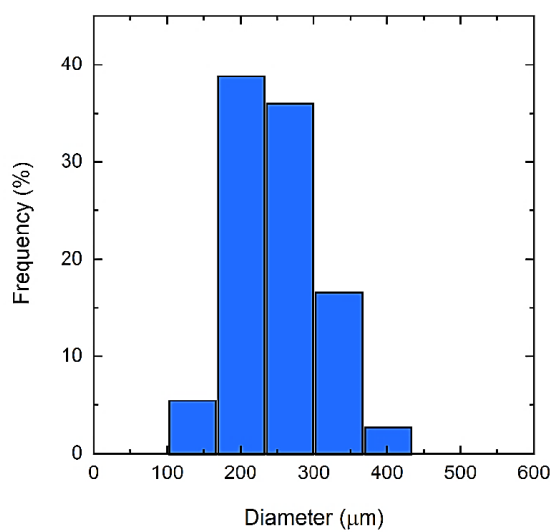

**Figure 4:** Size distribution for the alginate microcapsules of figure 3G.

The results showed a huge potential for microencapsulation to avoid ascorbic acid degradation in food products and, furthermore, can mask the acidic flavor of the compound [32].

Biocontrol using bacteriophages is a promising method with advantages compared to other methods. The phage is self-limiting, that is, it replicates only in the presence of the bacterium to which it has affinity; it also does not act in the selection of resistant microorganisms, can be administered in a single dose, is easy to use, is safe for humans, plants and animals, and phages are ubiquitous and diverse. However, bacteriophages can recover the replication in conditions of temperature abuse, or even exhibit a passive activity that cause reductions without the need for the bacteriophage to replicate and complete their life cycle. The ratio of host cells and phages denominated as "multiplicity of infection (MOI)" appears to be of great relevance to the successful application of this technology. Thus, it would be necessary to have lower concentrations than for other antimicrobial agents [33, 34].

Bacteriophages fit into the classification of natural antimicrobials, and their effectiveness in controlling bacterial pathogens in the agro-food industry has led to the development of different phage products already approved by the Food and Drug Administration (FDA) and the United States Department of Agriculture (USDA) [35]. Phage encapsulation is a technique with the possibility of reducing viral death during passage through the gastrointestinal tract, as well as an alternative to control the release of these particles, and microfluidic devices have been shown to be efficient in the encapsulation of phages as well as other drugs, cells, food ingredients, etc.

### 3.4. Confocal microscopy

By the confocal microscopy technique, the bacteriophages were observed in the interior of microcapsules in a homogeneous way, due to the presence of isothiocyanate (FITC) used to mark the phages. The results are demonstrated in Figure 3.

As shown in Figure 5, the labeling phage was present inside the microcapsules. These results show that the scanning confocal microscopy technique was efficient in demonstrating the presence of the phages. The retention efficiency test was also applied to the microcapsules with the labeled phage to confirm the results of microscopy. The marking methodology was proposed by Gitis and co-workers [36] and reproduced by Bakhshayeshi and contributors [22]. In both works, the conclusion was that the bacteriophages were marked in free amino groups on the lysine amino acid residues of their capsid proteins, using FITC or another coloring agent. Thus, this technique has a wide range of viruses for application, including mammal viruses and bacteriophages, due to the presence of the lysine amino acid in the capsid of these groups, allowing viral particles with different dimensions, shapes and physicochemical properties to be studied [22].

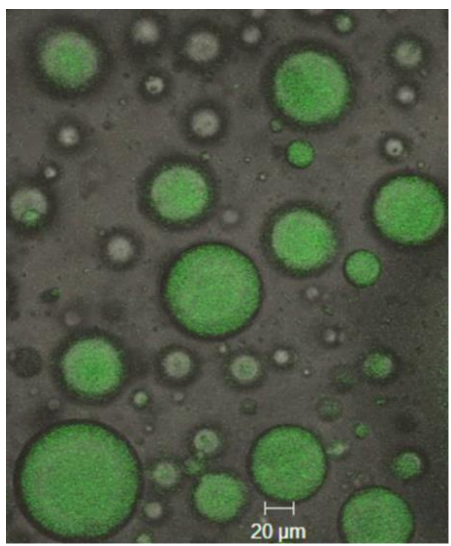

**Figure 5:** Scanning confocal microscopy of the alginate-Ca microcapsules containing the bacteriophages labeled by fluorescence. 200X magnification and fluorescence at a 488 nm excitation.

### 3.5. Stability of microcapsules in the propylene glycol gel and antimicrobial activity

The microcapsule stability results are presented in Table 1. For 21 days of storage at room temperature, the bacteriophages retained in the alginate microcapsules added to the gel remained viable, with no statistically different results from the control solution (alginate-Na 1.5% before injecting into the microfluidic device added to 106 PFU·mL$^{-1}$ solution of bacteriophages). This initial solution was used to produce the alginate-Ca microcapsules containing the bacteriophages and, after encapsulation, the capsules were added to the propylene glycol gel. The results demonstrate that the phages added in the alginate-Na solution were encapsulated in the alginate-Ca microcapsules and remained stable for 21 days after the microcapsules were added to the propylene glycol gel. This demonstrates that alginate is a good polymer to be used in encapsulation processes of materials, not affecting the stability and viability of the encapsulated material.

Furthermore, the alginate can be added to other polymer-based gels that are very stable, without affecting the structure of the alginate. The gel form presents as a stable suspension, being quite suitable for formulations of topical use. Gels are indicated for the delivery of water-soluble active substances and liposomes. They are most used in oily and mixed skins [37-39].

Using the bacteriophages in the polymeric matrix of calcium alginate and agar-agar to immobilize phages specific for Pseudomonas aeruginosa, Balcão and co-workers [40] demonstrated the maintenance of lytic viability of the viral particles after immobilization in the biopolymeric matrix, confirmed by the presence of lysis plaques in the bacterial growth medium. The authors produced the gel without any encapsulation device. Tang and co-workers [41] used alginate microspheres coated with whey protein for preparation of a dry powder containing the K bacteriophage for oral administration to remain viable until the intestine. Their study showed that the encapsulated phage remained viable in simulated gastric fluid at pH 2.5 after 2 hours of incubation. The encapsulated phage was completely released in the simulated intestinal juice within 2 hours. They concluded that the microspheres could be a vehicle to transport the K bacteriophage by oral administration to control *Staphylococcus aureus*.

The antimicrobial activity of the microencapsulated bacteriophages added to propylene glycol gel showed efficiency (not detectable on the surface) when applied to surfaces contaminated with *E. coli* O157:H7. The antimicrobial efficiency results were compared to an alcohol solution 70% applied to the same surface (not detectable in surface). Another surface without the application of any antimicrobial agents was used to compare the results. The bacterial growth on the surface without the application of antimicrobial agents was 4.92 log. It was observed that between the propylene glycol gel and the alcohol 70% there was no statistical difference in the antibacterial activity. Comparing with the control surface, the propylene glycol gel and the alcohol 70% were effective in sanitizing the surfaces. The initial inoculation of the surface was 104 CFU·mL$^{-1}$.

Using bacteriophages as antimicrobial agents in relation to other chemicals in the hygienization process of chicken skin contaminated with Salmonella sp., Hungaro et al. [42] observed that both bacteriophages and chemicals were efficient in reducing bacterial contamination. In this study, the phages were in solution and later sprayed over the chicken skin surface. Using cellulose acetate film incorporated with bacteriophages as an application in food packing, Gouvêa and co-workers [7] observed the antibacterial effect of the film containing the phages specific for Salmonella typhimurium in liquid medium and diffusion in solid medium, inhibiting bacterial growth. Bacteriophages incorporated in whey protein films applied in liquid medium and on food surfaces were released in the medium, causing antimicrobial effect after 6 h. The liberation was efficient both in liquid medium and on food surfaces [6].

**Table 1** Stability of the control alginate-Na before injecting into the microfluidic device added to the bacteriophage solution (A1) and of the alginate-Ca microcapsules containing the phages (A2) added to the propylene glycol gel during 21 days of storage at room temperature.

| | *Bacteriophage concentration (Log PFU·mL$^{-1}$)* | | | |
|---|---|---|---|---|
| **Samples** | **1 day** | **2 days** | **14 days** | **21 days** |
| A1 | 7.00 | 7.20 | 7.48 | 7.54 |
| A2 | 6.00 | 6.18 | 6.48 | 6.54 |

In our work, we can observe that the microencapsulated active principle (phage) remained stable during the 21 days of analysis of the gel; lytic activity of phage proved to be efficient after the application of the gel to the surface; and the microcapsules were ruptured via rubbing the gel, releasing phage to infect the bacterial cell. The bacteriophage, being an organic antimicrobial agent, shows different behavior from a chemical antimicrobial. As a virus, it needs the host to replicate, and therefore it must initially find the host in the environment, then infecting it and causing bacterial lysis (host death). Thus, the time of action of the bacteriophage is more dynamic. The chemical antimicrobial agent acts on the host immediately, eliminating it, but if there is a new infection, the antimicrobial chemical does not continue to act. In the case of bacteriophages, which are more dynamic, the process is better because it finds the host, infects it, replicates itself, causes lysis and releases new phage particles into the medium, restarting a new cycle. Within 24 h the multiplication process is still occurring because new virus particles are still being released to infect the host. Where the host is present, the bacteriophage is also able to restart a new cycle of replication and cell lysis.

## 4. Conclusion

The UFV-AREG1 bacteriophage was encapsulated in the calcium alginate matrix by the microfluidic technique using the capillary device with retention efficiency close to 90% of viral capsules in the capsules. The encapsulation was confirmed by confocal microscopy. Thereby, the capsules were applied in a propylene glycol gel, in which the bacteriophage remained viable for the 21 days of analysis. The antimicrobial gel application on contaminated surfaces was efficient to reduce bacterial contamination in relation to an antimicrobial chemical agent. Hence, it was possible to microencapsulate the phages in the alginate-Ca matrix and apply the capsules to antimicrobial gels to use as sanitizers in the food industry.

## Acknowledgements

The research group thanks the Nucleus of Microscopy and Microanalysis (NMM) of the Federal University of Viçosa (UFV) for the images. They are also grateful to the Laboratory of Microfluidic and Complex Fluids (LMFC) of the Department of Physics of UFV for the space, equipment and other contributions and to the Foundation for the Support of MG Research (FAPEMIG) for financial support in the project number APQ02232-12. No member of the group has any conflict of interest.

## References

[1] J.P. Kerry, M.N. O'Grady, S.A. Hogan, Past, current and potential utilisation of active andintelligent packaging systems for meat and muscle-based products: A review, Meat Science, 74(2006) 113-130.

[2] M. Ahmad, S. Benjakul, T. Prodpran, T.W. Agustini, Physico-mechanical and antimicrobialproperties of gelatin film from the skin of unicorn leatherjacket incorporated with essentialoils, Food Hydrocolloids, 28 (2012) 189-199.

[3] G. Lopez-Velasco, A. Tomas-Callejas, A.O. Sbodio, X. Pham, P. Wei, D. Diribsa, T.V. Suslow,Factors affecting cell population density during enrichment and subsequent moleculardetection of Salmonella enterica and Escherichia coli O157:H7 on lettuce contaminated duringfield production, Food Control, 54 (2015) 165-175.

[4] S.S. Shekarforoush, S. Basiri, H. Ebrahimnejad, S. Hosseinzadeh, Effect of chitosan onspoilage bacteria, Escherichia coli and Listeria monocytogenes in cured chicken meat,International Journal of Biological Macromolecules, 76 (2015) 303-309.

[5] J. Tsonos, D. Vandenheuvel, Y. Briers, H. De Greve, J.-P. Hernalsteens, R. Lavigne, Hurdles inbacteriophage therapy: Deconstructing the parameters, Veterinary Microbiology, 171 (2014)460-469.

[6] E. Vonasek, P. Le, N. Nitin, Encapsulation of bacteriophages in whey protein films forextended storage and release, Food Hydrocolloids, 37 (2014) 7-13.

[7] D.M. Gouvêa, R.C.S. Mendonça, M.L. Soto, R.S. Cruz, Acetate cellulose film withbacteriophages for potential antimicrobial use in food packaging, LWT - Food Science andTechnology, 63 (2015) 85-91.

[8] A. Choińska-Pulit, P. Mituła, P. Śliwka, W. Łaba, A. Skaradzińska, Bacteriophageencapsulation: Trends and potential applications, Trends in Food Science & Technology, 45(2015) 212-221.

[9] C.-X. Zhao, Multiphase flow microfluidics for the production of single or multiple emulsionsfor drug delivery, Advanced Drug Delivery Reviews, 65 (2013) 1420-1446.

[10] J.N. Schianti, N.N.P. Cerize, A.M. Oliveira, S. Derenzo, M.R. Góngora-Rubio, Scaling up ofRifampicin Nanoprecipitation Process in Microfluidic Devices, Progress in Nanotechnology andNanomaterials 2(2013) 101-107.

[11] M.J. Madou, Fundamentals of Microfabrication: The Science of Miniaturization, SecondEdition, Taylor & Francis2002.

[12] A.S. Utada, E. Lorenceau, D.R. Link, P.D. Kaplan, H.A. Stone, D.A. Weitz, Monodispersedouble emulsions generated from a microcapillary device, Science, 308 (2005) 537-541.

[13] N.E. Simpson, C.L. Stabler, C.P. Simpson, A. Sambanis, I. Constantinidis, The role of theCaCl2–guluronic acid interaction on alginate encapsulated βTC3 cells, Biomaterials, 25 (2004)2603-2610.

[14] V. Nedovic, A. Kalusevic, V. Manojlovic, S. Levic, B. Bugarski, An overview of encapsulationtechnologies for food applications, Procedia Food Science, 1 (2011) 1806-1815.

[15] S. Martins, B. Sarmento, E.B. Souto, D.C. Ferreira, Insulin-loaded alginate microspheres fororal delivery – Effect of polysaccharide reinforcement on physicochemical properties andrelease profile, Carbohydrate Polymers, 69 (2007) 725-731.

[16] G. Lawrie, I. Keen, B. Drew, A. Chandler-Temple, L. Rintoul, P. Fredericks, L. Grøndahl,Interactions between Alginate and Chitosan Biopolymers Characterized Using FTIR and XPS,Biomacromolecules, 8 (2007) 2533-2541.18

[17] W.R. Gombotz, S. Wee, Protein release from alginate matrices,

Advanced Drug DeliveryReviews, 31 (1998) 267-285.

[18] M. Liu, X.-T. Sun, C.-G. Yang, Z.-R. Xu, On-chip preparation of calcium alginate particlesbased on droplet templates formed by using a centrifugal microfluidic technique, Journal ofColloid and Interface Science, 466 (2016) 20-27.

[19] M.E.S. Lopez, L.S. Batalha, P.M.P. Vidigal, L.A.A. Albino, D.M.G. Boggione, M.T.P. Gontijo,D.M.S. Bazolli, R.C.S. Mendonça, Escherichia phage UFV-AREG1, complete genome, GenBank,2016.

[20] X.D. Liu, W.Y. Yu, Y. Zhang, W.M. Xue, W.T. Yu, Y. Xiong, X.J. Ma, Y. Chen, Q. Yuan,Characterization of structure and diffusion behaviour of Ca-alginate beads prepared withexternal or internal calcium sources, Journal of Microencapsulation, 19 (2002) 775-782.

[21] C. Gomes, R.G. Moreira, E. Castell-Perez, Microencapsulated antimicrobial compounds asa means to enhance electron beam irradiation treatment for inactivation of pathogens onfresh spinach leaves, J Food Sci, 76 (2011) E479-488.

[22] M. Bakhshayeshi, N. Jackson, R. Kuriyel, A. Mehta, R. van Reis, A.L. Zydney, Use of confocalscanning laser microscopy to study virus retention during virus filtration, Journal of MembraneScience, 379 (2011) 260-267.

[23] A.N. Martin, P. Bustamante, Physical Pharmacy: Physical Chemical Principles in thePharmaceutical Sciences, Lea & Febiger1993.

[24] Y. Hu, Q. Wang, J. Wang, J. Zhu, H. Wang, Y. Yang, Shape controllable microgel particlesprepared by microfluidic combining external ionic crosslinking, Biomicrofluidics, 6 (2012)026502-026502-026509.

[25] Y. Hu, G. Azadi, A.M. Ardekani, Microfluidic fabrication of shape-tunable alginatemicrogels: effect of size and impact velocity, Carbohydr Polym, 120 (2015) 38-45.

[26] C.-H. Yeh, Q. Zhao, S.-J. Lee, Y.-C. Lin, Using a T-junction microfluidic chip formonodisperse calcium alginate microparticles and encapsulation of nanoparticles, Sensors andActuators A: Physical, 151 (2009) 231-236.

[27] T. Nisisako, T. Torii, T. Higuchi, Droplet formation in a microchannel network, Lab on aChip, 2 (2002) 24-26.

[28] K.-S. Huang, Y.-S. Lin, C.-H. Yang, C.-W. Tsai, M.-Y. Hsu, In situ synthesis of twinmonodispersed alginate microparticles, Soft Matter, 7 (2011) 6713-6718.

[29] P.-W. Ren, X.-J. Ju, R. Xie, L.-Y. Chu, Monodisperse alginate microcapsules with oil coregenerated from a microfluidic device, Journal of Colloid and Interface Science, 343 (2010) 392-395.

[30] S.-B. Huang, M.-H. Wu, G.-B. Lee, Microfluidic device utilizing pneumatic micro-vibratorsto generate alginate microbeads for microencapsulation of cells, Sensors and Actuators B:Chemical, 147 (2010) 755-764.

[31] L. Liu, F. Wu, X.-J. Ju, R. Xie, W. Wang, C.H. Niu, L.-Y. Chu, Preparation of monodispersecalcium alginate microcapsules via internal gelation in microfluidic-generated doubleemulsions, Journal of Colloid and Interface Science, 404 (2013) 85-90.

[32] T.A. Comunian, A. Abbaspourrad, C.S. Favaro-Trindade, D.A. Weitz, Fabrication of solidlipid microcapsules containing ascorbic acid using a microfluidic technique, Food Chemistry,152 (2014) 271-275.

[33] M.R.J. Clokie, A.M. Kropinski, Bacteriophages: Methods and Protocols, Volume 1:Isolation, Characterization, and Interactions, Humana Press2009.

[34] D.M. Knipe, P. Howley, Fields Virology, Wolters Kluwer Health2013.

[35] S.M. Sillankorva, H. Oliveira, J. Azeredo, Bacteriophages and Their Role in Food Safety,International Journal of Microbiology, 2012 (2012) 13.

[36] V. Gitis, A. Adin, A. Nasser, J. Gun, O. Lev, Fluorescent dye labeled bacteriophages—a newtracer for the investigation of viral transport in porous media: 1. Introduction andcharacterization, Water Research, 36 (2002) 4227-4234.19

[37] G. Leonardi, P. Maia Campos, Estabilidade de formulações cosméticas, InternationalJournal of Pharmaceutical Compounding, 3 (2001) 154-156.

[38] G.R. Leonardi, L. Matheus, Cosmetologia aplicada, São Paulo (SP): Medfarma, (2004).

[39] N.M. Corrêa, F.B.C. Júnior, R.F. Ignácio, G.R. Leonardi, Avaliação do comportamentoreológico de diferentes géis hidrofílicos, Revista Brasileira de Ciências Farmacêuticas, 41(2005).

[40] V.M. Balcão, A.R. Moreira, C.G. Moutinho, M.V. Chaud, M. Tubino, M.M.D.C. Vila,Structural and functional stabilization of phage particles in carbohydrate matrices for bacterialbiosensing, Enzyme and Microbial Technology, 53 (2013) 55-69.

[41] Z. Tang, X. Huang, P.M. Sabour, J.R. Chambers, Q. Wang, Preparation and characterizationof dry powder bacteriophage K for intestinal delivery through oral administration, LWT - FoodScience and Technology, 60 (2015) 263-270.

[42] H.M. Hungaro, R.C.S. Mendonça, D.M. Gouvêa, M.C.D. Vanetti, C.L.d.O. Pinto, Use ofbacteriophages to reduce Salmonella in chicken skin in comparison with chemical agents, FoodResearch International, 52 (2013) 75-81.